\pdfoutput=1
\documentclass[fleqn,10pt]{wlscirep}
\usepackage[utf8]{inputenc}
\usepackage[T1]{fontenc}
\title{Multibit neural inference in a $N$-ary crossbar architecture}

\author[1,*]{Anatole Moureaux}
\author[1]{Anthony Lopes Temporao}
\author[1]{Flavio Abreu Araujo}
\affil[1]{Université catholique de Louvain, Institute of Condensed Matter and Nanosciences, Louvain-la-Neuve, 1348, Belgium}
\affil[*]{anatole.moureaux@uclouvain.be}


\begin{abstract}
In-memory computing (IMC) is a paradigm that enables neural network inference by computing analog matrix-vector multiplications (MVM) directly in memory crossbar arrays, with the potential for energy efficiency gains over conventional von Neumann architectures.
In this work we present a simulation framework for $N$-ary crossbar architectures that retrieves MVM results with minimal implementation assumptions. 
The XOR and MNIST classification tasks were successfully inferred using a simulated crossbar array of (4 $\times$ 4) 4-states magnetic tunnel junctions (MTJ).
MNIST accuracy reached 93.56\% (vs. 97.56\% software baseline). 
PCA dimensionality reduction was shown to drastically lower the number of required operations and improve the software baseline, for only a modest reduction in crossbar inference accuracy.
We identified weight quantization as the primary error source, and studied its impact alongside systematic non-idealities and random noise.
We find that cell-specific random noise is less detrimental than systematic errors due to averaging across the array.
Finally, we demonstrate an optimal number of states per cell that balances quantization error against resistance state resolution to minimize total MVM error.
\end{abstract}
\begin{document}

\flushbottom
\maketitle
%
%
\thispagestyle{empty}
 
\section*{Introduction}

Computing systems based on the von Neumann architecture face fundamental efficiency limitations. 
The first one is a structural constraint known as the von Neumann bottleneck, caused by the physical separation between the memory (RAM) and the processing unit (CPU).
This separation requires continuous data movement across a shared bus, resulting in significant energy consumption~\cite{Zou2021}.
The other limitation is known as the memory wall and arises from the increasing gap between the speed of processors and memory access times~\cite{Wulf1995}.
These limitations have become even more concerning recently with the growth of artificial intelligence (AI) models and of data scales, leading to significant energy consumption and processing latency. 
This observation drove the demand for new computing systems targeting computational performance and energy efficiency~\cite{BigData, EnergyForecast}.
However, even with specialized accelerators such as GPUs and CMOS-based application specific integrated circuits (ASICs), limitations persist due to constrained on-chip memory, leakage currents, and limited data parallelism~\cite{Chen2016, Sim2016, Desoli2017, Moons2017}.
These challenges have become critical barriers to further progress, driving the need for a fundamental rethinking of computing paradigms.

In-memory computing (IMC) is a promising paradigm that integrates data storage and computation within the same physical units, thereby addressing the limitations mentioned previously~\cite{Sebastian2020}.
In particular, IMC enables the highly efficient inference of deep neural networks (DNN) by doing analog matrix-vector multiplications (MVM) inside the memory.
MVMs are the core operations in neural networks, multiplying input vectors $\mathbf{x}$ representing data by weight matrices $\mathbf{W}$ learned during the training phase to propagate the resulting output vector $\mathbf{y}$ through the network layers: 
\begin{equation}
    \mathbf{y} = \mathbf{W} \mathbf{x}
    \label{eq:mvm_eq}
\end{equation}

The most suited hardware for this operation are crossbar arrays of cells with distinct programmable conductance levels.
The input is injected in the array as a voltage vector, and is multiplied by the conductance of each cell in a given row.
The currents produced in each cell owing to Ohm's law are then accumulated along each column to produce the output, accordingly to Kirchhoff's law.
Several types of non-volatile memories (NVM) have been explored for designing cells in such crossbar arrays, including resistive random-access memory (RRAM), phase-change memory (PCM), and flash memory~\cite{Sebastian2020, Ikegawa2020}.
Among emerging non-volatile memory technologies, magnetoresistive random-access memories (MRAM) stand out as particularly promising candidate for IMC due to their intrinsic non-volatility, high endurance, energy efficiency, and compatibility with CMOS technology~\cite{Fong2015}.

Despite the advantages of analog computation in crossbar arrays, effective integration of IMC within large-scale AI systems necessitates algorithms capable of translating the array's analog input/output signals into the digital signals required by standard AI algorithms.
Indeed, input vectors of data require digital-to-analog conversion in order to be injected in the crossbar array. Similarly, the output of analog MVM operations must often be converted back into digital signals for reliable communication with subsequent compute units in the neural network like activation functions and biasing.
These conversions are facilitated by digital-to-analog converters (DAC) and analog-to-digital converters (ADC), which introduce significant challenges: they not only contribute disproportionately to the system energy and area overhead, but also impact computational accuracy due to quantization errors and analog non-idealities~\cite{Yu2018, Chakraborty2020}.
Therefore, specialized algorithm-hardware co-design is essential for optimizing these interfaces, managing precision requirements, mitigating signal degradation, and compensating for device-level non-idealities.

In this regard, a wide variety of algorithm-hardware co-design approaches have emerged, leveraging recent quantized neural network (QNN) research~\cite{Wang2018, Han2015, Hubara2018, Li2016, Zhou2016} to fit with the low-bit cells currently used.
For instance, memristive QNNs, spin-transfer torque binary neural networks (STT-BNNs), and IMC based on spin-orbit torque magnetoresistive random access memories (SOT-MRAM) frameworks exploit binary and ternary weight schemes for robust in-situ learning and inference with reduced digital interfacing~\cite{Zhang2019memristive, Pham2022stt, Jung2022, Doevenspeck2020sot}.
Stochastic magnetic tunnel junctions (MTJ) harness intrinsic probabilistic switching for QNN training without the random number generation (RNG) overhead~\cite{Greenberg2021}, while FeFET arrays encode multiply-and-accumulate (MAC) results in time-domain activation delays, eliminating full analog-digital multipliers~\cite{Soliman2023}.
$\text{MoS}_2$ transistor-based 2T-1C cells perform analog MAC operations with recalibrated weights to ensure linearity under multilevel storage~\cite{Wang2021memory}, and spintronic resonators use frequency-coded weights to classify RF signals without requiring digitization~\cite{Leroux2021}.
Domain-wall spin-orbit torque (DW-SOT) devices combine logic and storage for multistate Boolean operations~\cite{Lin2022}, while multi-state magnetic tunnel junctions enable quantized analog synapses with low latency and energy~\cite{Rzeszut2022, Das2020}.

In this paper we propose a simulation-based framework for running multibit AI inference on a $N$-ary crossbar architecture \textit{i.e.}, whose cells present more than 2 distinct states.
We first present an overview of the typical $N$-ary crossbar architecture, followed by our solution for the retrieval of the actual MVM result based solely on the input signal and the measured output signal.
A key advantage of our approach is that it allows to use the crossbar array in its simplest form as a standalone AI coprocessor without affecting its internal architecture nor the rest of the inference pipeline.
Furthermore, it directly multiplies analog input vectors by the optimal quantized version of the weights matrix, without any additional hypotheses on the physics of the crossbar array cells, making it generalizable to any kind of $N$-ary crossbar array implementation without loss of generality.
In our case, we showcase the use of the framework within two simple benchmarking tasks, using digital twins of the magnetic tunnel junctions presented in Refs.~\cite{Das2020, Das2020stabilization}, which display 4 distinct resistance states, as a starting point (as better metrics probably exist now). These MTJs will be referred to as multistate magnetic tunnel junctions (M²TJs) in the rest of the paper, so as proposed in Ref.~\cite{MsAI}. 

A 4-by-4 crossbar arrays of 4-states M²TJs was simulated to infer a neural network that was trained to learn the XOR function.
The results obtained with the simulated crossbar array are equivalent to the ones obtained with the full-digital neural network, showing that the intrinsic properties of the model learned during the training phase are preserved within the crossbar array-based inference process.
Then, a network trained to classify the MNIST handwritten digits dataset was inferred using the simulated crossbar array. 
Accuracy levels (93.56\%) close to the software baseline (97.56\%) were observed. 
We show that the primary source of error responsible for this decrease in accuracy is the quantization of the initially full-precision weights into the limited number of resistance levels of the crossbar array cells, highlighting the importance of multibit implementations.   
In order to reduce the number of operations required on the limited size crossbar array, the input dimensionality is reduced through principal components analysis (PCA).
We show that this dimensionality reduction drastically lowers the number of required sub-operations and improves the software baseline, for only a relatively small reduction in crossbar inference accuracy, making it interesting for deploying larger networks on small crossbar arrays.
Two other sources of error are finally studied: systematic non-idealities affecting all the cells of the crossbar array identically, and random noise affecting each cell independently due to thermal fluctuations and device-to-device variations.
Both error sources result in an error in the MVM result that increases linearly with the amplitude of the source.
However, we show that cell-specific errors have a less detrimental impact on the inference accuracy than systematic errors of similar amplitude, due to the beneficial averaging effect across the whole array.
Finally, we observe the existence of an optimal number of states per cell that minimizes the error in the MVMs by balancing weights precision and noise robustness. 
We demonstrate through simulation that this number evolves with respect to the level of noise affecting the crossbar array, and conclude that smaller numbers of states are preferable in noisier environments, a crucial information for the future development of experimental validation platforms.

\begin{figure}[ht]
    \centering
    \includegraphics[width=.5\linewidth]{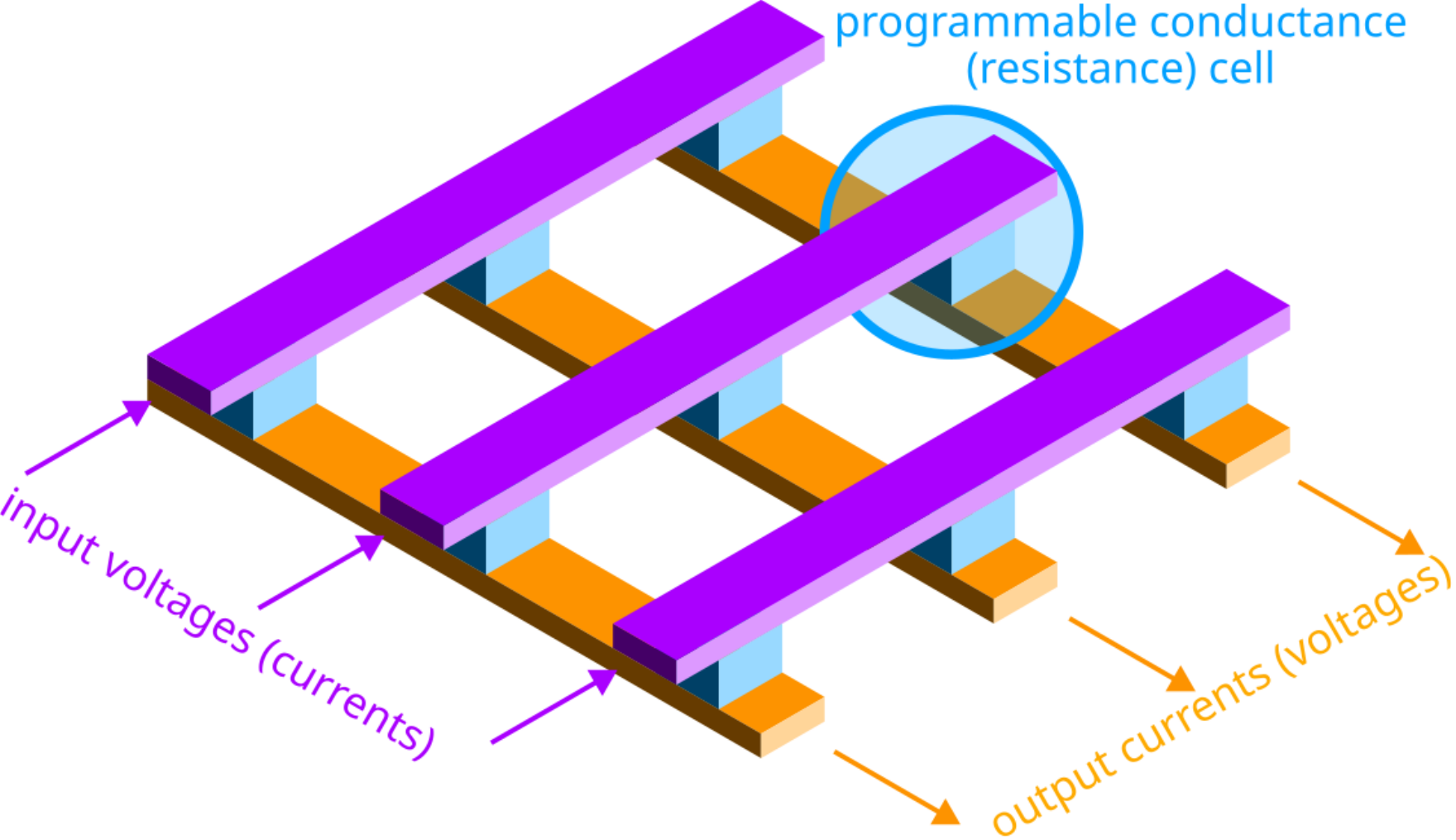}
    \caption{A typical crossbar architecture of ($3 \times 3$) memristive cells.}\label{fig:crossbar}
\end{figure}

The typical crossbar architecture (see Fig.~\ref{fig:crossbar}) is composed of a grid of cells, each of them having a conductance $G$ that can be programmed to a set of $N$ distinct levels.
According to Ohm's law, the current through each cell is given by Eq.~\ref{eq:mac_eq} where $G_{m n}$ is the conductance of the cell at row $m$ and column $n$, and $V_n$ is the input voltage applied to column $n$.
\begin{equation}
    I_{m n} = G_{m n} V_n
    \label{eq:mac_eq}
\end{equation}
By Kirchhoff's law, the total output current at row $m$ is the sum of the currents from all crosspoints in that row (Eq.~\ref{eq:mac_eq2}).
\begin{equation}
    I_m = \sum_n I_{m n}
    \label{eq:mac_eq2}
\end{equation}
This operation naturally implements an analog MAC operation, where the conductance levels encode the weights and the input voltages represent the input vector.
The converse configuration is also possible: the input and output vectors would respectively be encoded as currents and voltages vectors, and the weights would be encoded as resistance values as presented in Eq.~\ref{eq:mac_eq1bis}.
The voltages produced across each cell would then be summed along each row to produce the output voltage vector (Eq.~\ref{eq:mac_eq2bis})
The choice of the configuration has to be made in consideration of the devices selected as memristive cells, so that a set of distinct and reproducible conductance or resistance levels is accessible.
\begin{equation}
    V_{m n} = R_{m n} I_n
    \label{eq:mac_eq1bis}
\end{equation}
\begin{equation}
    V_m = \sum_n V_{m n}
    \label{eq:mac_eq2bis}
\end{equation}
Summing currents or voltages are however not the only options. For example, Ref.~\cite{Jung2022} introduces resistance summation, which overcomes the low-resistance issue of MRAM devices that leads to extensive power consumption by combining several memristive devices in a single cell.
This work focuses on the M²TJs presented in Refs.~\cite{Das2020, Das2020stabilization}, whose resistance is a function of the device magnetization state.
By controlling the magnetization of the free layer, which is made of $k$ crossed ferromagnetic ellipses, each cell can access $k$, $\text{2}^k$, or $\text{2}^{\text{2} k}$ distinct magnetic configurations depending on the method used for writing the states~\cite{Das2020stabilization}.
The resistance states exhibited by these devices are well separated and reproducible~\cite{Das2020}, which is the key requirement for efficient $N$-ary MAC operations.
The framework does not require these levels to be equidistant. 
As detailed in the Methods, an array-aware quantization step assigns to the quantized weights the same relative spacing as the measured resistance levels, so that an exact linear state-to-resistance mapping is preserved for any arbitrary level distribution.
Despite being focused on this specific device, the framework presented in this paper is therefore generalizable to any crossbar array made of cells featuring distinct and reproducible resistance or conductance levels in an increasing order.

\section*{Results}

\subsection*{Toy example: XOR function}
We first ensure that the basic properties of a trained neural network are preserved within the crossbar array-based inference process by considering the XOR approximation task.
Exclusive OR (XOR) is a binary function defined as $(x_1 \land \lnot x_2) \lor (\lnot x_1 \land x_2)$
Its truth table is presented in Table~\ref{tab:xor_truth_table}.
It is is nonlinearly separable, meaning that the inputs $(x_1, x_2)$ for which the output $y$ is 0 cannot be separated with a single line in the input space from the inputs for which the output is 1, making it useful to assess the basic nonlinear separation performance of neural networks.
\begin{table}[ht]
    \centering
    \begin{tabular}{|c|c|c|}
        \hline
         $\mathbf{x_1}$ & $\mathbf{x_2}$ & $\mathbf{y = XOR(x_1, x_2)}$\\
         \hline
         $0$ & $0$ & $0$\\
         \hline
         $0$ & $1$ & $1$\\
         \hline
         $1$ & $0$ & $1$\\
         \hline
         $1$ & $1$ & $0$\\
         \hline
    \end{tabular}
    \caption{\label{tab:xor_truth_table} Truth table of the XOR function.}
\end{table}
We first train a simple artificial neural network (ANN) to learn the XOR function.
The ANN is composed of a 2-neurons input layer, a 2-neurons hidden layer, and a 1-neuron output layer, all activated with the sigmoid function.
The ANN was trained with the Adam optimizer and a learning rate of 0.01 for 2000 epochs, using the binary cross-entropy loss function.
After training, the ANN achieved a training accuracy of 100\% for the input values of Table~\ref{tab:xor_truth_table}.
We illustrate the use of our framework with a theoretical crossbar array of  (2 $\times$ 2) M²TJs presenting the 4 distinct resistance levels $R_i$ reported in Ref.~\cite{Das2020} (Table~\ref{tab:das_resistances}).
As these measured levels are not equidistant, our framework is made \textit{array-aware}: the quantization states $A_i$ are chosen so that their relative spacings match those of the measured resistance levels, which guarantees that the state-to-resistance mapping of Eq.~\ref{eq:r_levels} is perfectly linear (see Methods).
To simplify the example, we also assume that the cells do not present any device-to-device variations and are not subject to noise.
Different sources of error will be studied in a following section.
\begin{table}
    \centering
    \begin{tabular}{|c|c|}
        \hline
        \textbf{State} & \textbf{Resistance value ($\Omega$)}\\
        \hline
        State 1 & $9357$\\
        \hline
        State 2 & $9645$\\
        \hline
        State 3 & $9741$\\
        \hline
        State 4 & $9900$\\
        \hline
    \end{tabular}
    \caption{\label{tab:das_resistances} Measured resistance levels of the 4-states M²TJs reported in Ref.~\cite{Das2020} and used in the toy example. The levels are not equidistant.}
\end{table}
The input vector $\mathbf{x} = (x_1, x_2)$ is linearly encoded as a current vector $\mathbf{I} = (I_1, I_2)$, where $I_1$ and $I_2$ are the currents injected into the first and second columns of the crossbar array, respectively.
The scaling factors and offsets are chosen to bound the input signal between 0 mA and 0.5 mA.
\begin{equation}
    I_n = x_n/2 \text{ mA, with } x_n \in [0, 1]
    \label{eq:scalings_I}
\end{equation}
The weight matrix $\textbf{W}$ of the first layer was quantized into a matrix $\textbf{A}$ whose values belong to the set ${A_1, A_2, A_3, A_4}$ using the quantization method of Eq.~\ref{eq:opti}. 
Note that in this case, the quantization process yields only two distinct values (Eq.~\ref{eq:eq_quant}).
\begin{equation}
    \textbf{W}: \begin{pmatrix}11.97 & 12.06\\ 8.57& 8.58\end{pmatrix} \rightarrow \textbf{A}: \begin{pmatrix}12.02 & 12.02\\ 8.57 & 8.57\end{pmatrix}
    \label{eq:eq_quant}
\end{equation}
The values of $\textbf{A}$ are then linearly mapped to the measured resistance levels of Table~\ref{tab:das_resistances} through the linear relation obtained by least-squares fitting (Eq.~\ref{eq:scalings_R}).
Because the states $A_i$ share the relative spacing of the resistance levels, this fit is exact and introduces no residual mapping error.
\begin{equation}
    R_{m n} = 157.84\, A_{m n} + 8003.61 \quad \Omega
    \label{eq:scalings_R}
\end{equation}
Finally, the scaling factors from Eq.~\ref{eq:scalings_I} and Eq.~\ref{eq:scalings_R} are used to solve Eq.~\ref{eq:eq_VB} and simulate the output of the crossbar array.
The procedure is then repeated for carrying out a second MVM, involved in the output layer.
The results summarized in Table~\ref{tab:xor_summary} demonstrate that the crossbar array inference effectively yields the same results as the software ground truth, numerical inaccuracies apart.
\begin{table}[ht]
    \centering
    \begin{tabular}{|c|c|c|c|c|}
        \hline
        $\mathbf{x_1}$ & $\mathbf{x_2}$ & $\mathbf{y}$ & $\mathbf{\hat{y}_\text{CPU}}$ & $\mathbf{\hat{y}_\text{crossbar}}$\\
        \hline
        $0$ & $0$ & $0$ & $0.$ & $0.000$ \\
        \hline
        $0$ & $1$ & $1$ & $1.$ & $0.998$ \\
        \hline
        $1$ & $0$ & $1$ & $1.$ & $0.998$ \\
        \hline
        $1$ & $1$ & $0$ & $0.$ & $0.000$ \\
        \hline
    \end{tabular}
    \caption{\label{tab:xor_summary} Comparison of the XOR output using a CPU and a crossbar array for the inference phase.}
\end{table}

\begin{figure}[ht]
\centering
\includegraphics[width=.8\linewidth]{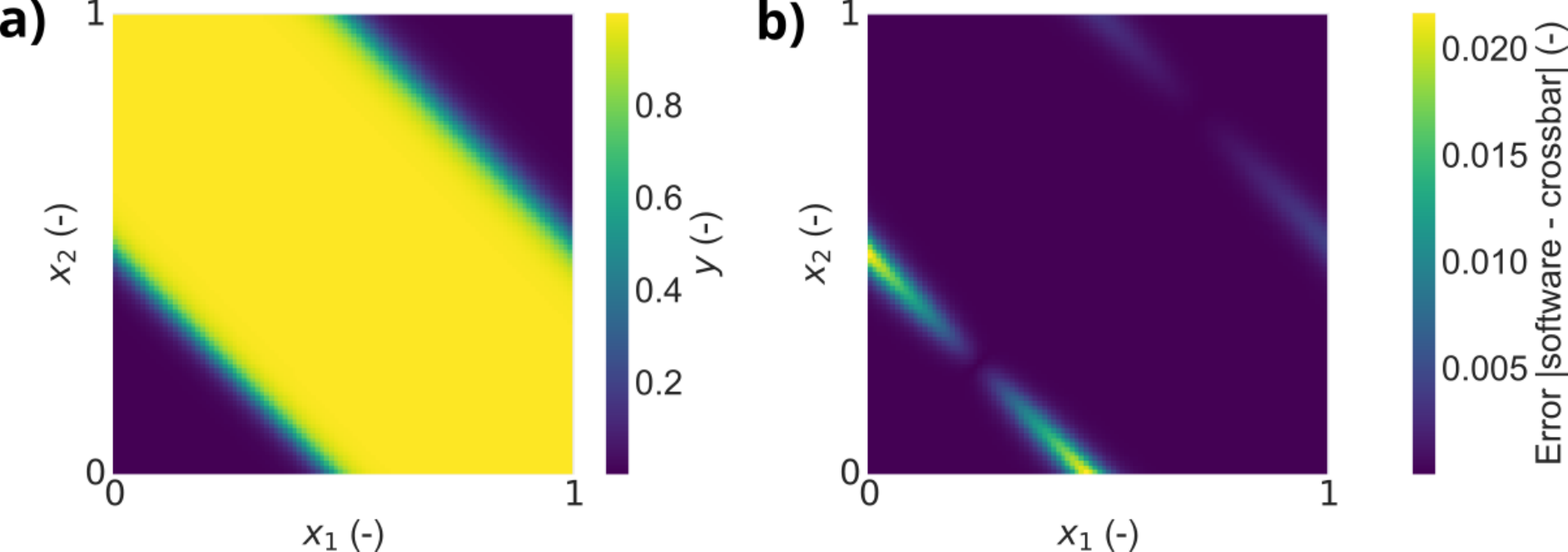}
\caption{\textbf{a)} Output of the software-inferred network. \textbf{b)} Absolute difference between the software output (ground truth) and the crossbar array output.}\label{fig:comp_xor}
\end{figure}

In order to get a better insight of the crossbar array inference process accuracy, the software output (Figure~\ref{fig:comp_xor}~(a)) was compared with the crossbar array output for any input in [0, 1] $\times$ [0, 1]. 
So far, the only source of error considered is the quantization of the weights into 4 values. 
It can be seen in Figure~\ref{fig:comp_xor}~(b) that this quantization error is mainly located around the boundaries between the two classes (in the lower left and upper right corners of the map) \textit{i.e.}, where $y$ is close to 0.5.
On the other hand, the error in the crossbar inference output is negligible for the actual inputs of interest of the XOR function (located in the corners of the input space), meaning that the crossbar array inference process preserves the learned properties of the model.

\subsection*{MNIST classification}
The same study was carried out for the MNIST handwritten digits classification task in order to assess the performance of the crossbar array inference on a more complex case.
A neural network was trained to classify the MNIST dataset, with an input layer of 784 neurons accounting for the (28 $\times$ 28) pixels of the images, a hidden layer of 128 neurons activated with the reLU function, and an output layer of 10 neurons activated with the sigmoid function, each neuron corresponding to a digit from 0 to 9. 
The final answer is given by the neuron of the output layer presenting the highest value.
After training, the accuracy of the model reached \textbf{97.56\%} on the 10000 samples of the testing set.
The same procedure as in the XOR case was used to infer the network using a crossbar array of (4 $\times$ 4) 4-states M²TJs.
The weights were quantized through the same array-aware scheme as before (see Methods), so that they can be matched to the measured resistance levels of Table~\ref{tab:das_resistances}.
To do so, the weights matrix $\textbf{A}$ was split into several submatrices of shape (4 $\times$ 4) that were sequentially written into the crossbar array to compute parts of the output vector $\tilde{\textbf{y}}$, a method referred to as \textit{multiplexing} and presented in Ref.~\cite{Jung2022}.
The input must also be modified several times to carry out the whole MVM with the crossbar.
The total number of sub-operations required to perform the MVM is given by 
$\lceil m/4 \rceil \times \lceil n/4 \rceil$ where $m$ and $n$ are respectively the number of rows and columns in the matrix $\textbf{A}$, totalizing 6272 sub-operations in our case for only carrying out the first MVM.\@
These numerous intermediary operations are a consequence of the limited size of the simulated crossbar array, and are expected to be reduced in the future with the development of larger crossbar arrays. 
The accuracy of the crossbar array inference process reached \textbf{93.56\%}, marking a drop of \textbf{4.00\%} accuracy from the software baseline due to the quantization error. 

While scaling aspects such as weight storage and tiling belong to the architectural design of the system and are strongly dependent on the specific implementation of the crossbar array and its integration, some techniques such as dimensionality reduction can be used to reduce the size of the weights matrices and hence reduce the number of required sub-operations without any loss of generality.
We applied principal component analysis (PCA)~\cite{Ringner2008} to reduce the input data dimensionality from 784 to 87 without significant loss of information (90\% of data variance preserved) before training the model. 
Only the images of the training set were used to compute the principal components, which were extracted from both training and testing sets before respectively training and testing the model.
This allowed to reduce the number of required sub-operations down to 704, which represents a 88.69\% decrease compared to the non-reduced case.
PCA also allows to reduce the amount of noise naturally present in the input data, which helps improving the performance of the model.
For that reason, the software accuracy of the model after PCA dimensionality reduction reached \textbf{98.03\%}. 
On the other hand, the inference of the simplified network with the simulated crossbar array led to an accuracy of \textbf{90.93\%}, marking a drop of \textbf{7.10\%} from the software baseline.
For a crossbar array of limited size, the number of sub-operations is the main practical bottleneck. 
This dimensionality reduction is therefore highly beneficial: it lowers the number of required sub-operations by $88.69\%$ while simultaneously denoising the input data and raising the software baseline from $97.56\%$ to $98.03\%$.
The crossbar inference of the simplified network reaches $90.93\%$, a modest $2.63\%$ below the full-network crossbar accuracy ($93.56\%$) for nearly an order of magnitude fewer operations.
The slightly larger gap with the software baseline ($7.10\%$ against $4.00\%$) comes from the fact that the weight distribution reshaped with PCA is quantized with a larger relative error.
However, this residual cost is largely compensated by the drastic reduction in operation count with limited-size arrays, making PCA a practical solution for deploying larger networks on small crossbar arrays.
The scaling of the error in larger crossbar arrays or multiplexed arrays is assessed in the last section of this paper.

\subsection*{Error assessment}
The error studies presented in this section characterize the framework independently of the devices used.
The studies are therefore based on a generic crossbar array with $N$ equidistant resistance levels rather than with specific non-equidistant levels like in the toy example, allowing the number of states $N$ to be swept freely.
The non-idealities $\sigma_\text{NL}$ and $\sigma_\perp$ are modeled as deviations from this equidistant reference.

\subsubsection*{Quantization error}
\begin{figure}[ht]
\centering
\includegraphics[width=.5\linewidth]{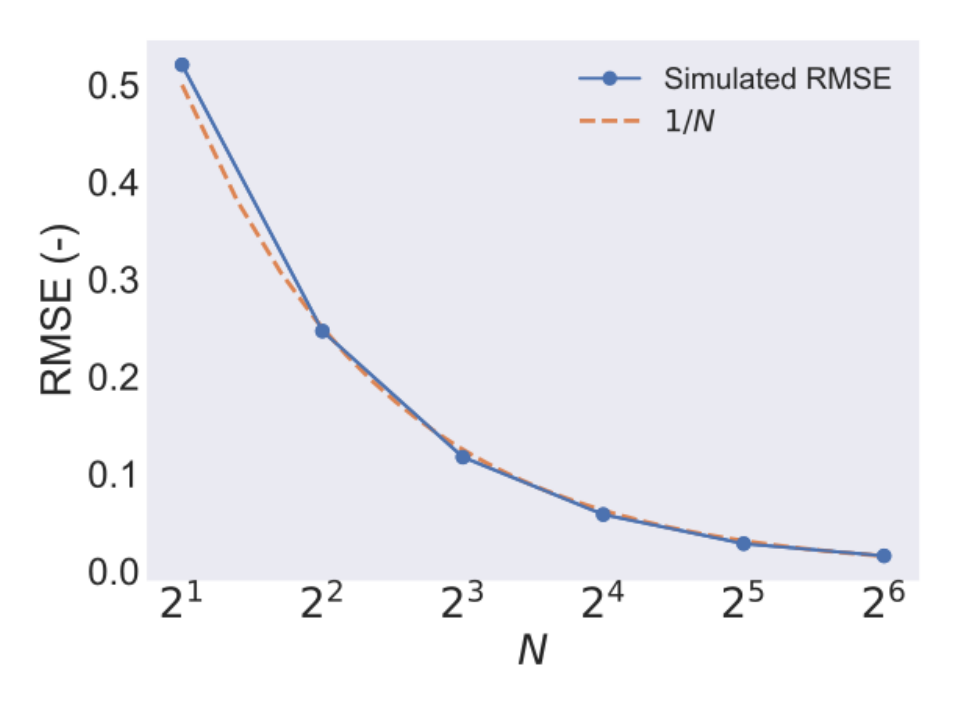}
\caption{Quantization error in randomized MVM results with respect to the number of states $N$ in each cell.}
\label{fig:fig_quant_error}
\end{figure}

The quantization of full-precision weights into 2-bits values is the only source of error considered so far in the pipeline.
To assess its impact on the accuracy of the crossbar array-based MVMs, we simulated MVMs with random (4 $\times$ 4) matrices with values drawn from $\mathcal{N}(\mu = 0, \sigma = 1)$ and input vectors of 4 values drawn from $\mathcal{U}(0, 1)$.
The root-mean-squared error (RMSE) between the software ground truth and the result obtained with the simulated crossbar array was then computed. 
The RMSE was averaged over 2000 MVMs, and the number of states per cell $N$ was swept from 1 to 64, marking 6-bits precision weights.
While such large numbers of resistance levels per cell are not yet accessible with current memristive technologies, an increase of $N$ can be expected in the future with the development of multistate memristive devices, with some recent works reporting up to 16 states per cell~\cite{Das2020stabilization}.
Without much surprise, the quantization RMSE follows a clear $1/N$ evolution due to the increasing bit precision of the weights as depicted in Figure~\ref{fig:fig_quant_error}.
While the development of multistate memristive devices with a higher number of states per cell is still a challenge, this result highlights the significant performance gain allowed by the use of multistate devices compared to binary ones, even if $N$ is only 3 or 4.

\subsubsection*{Input/output quantization}
In a physical crossbar array setup, input signals are generated by a wave generator whose digital-to-analog converter (DAC) introduces finite precision, and output voltages are read by an oscilloscope whose analog-to-digital converter (ADC) similarly introduces quantization.
To assess the contribution of this interface quantization to the total MVM error, we repeated the randomized MVM simulation of the previous section while quantizing both the input current vector $\mathbf{I}$ and the output voltage vector $\mathbf{V}$ on $b$ bits, and sweeping $b$ from 4 to 16.
The input is quantized over the range $[I_\text{min}, I_\text{max}] = [0, 0.5]$ mA, and the output over the theoretical full-scale range $[0, n R_\text{max} I_\text{max}]$ with $n$ the number of columns in the array.

\begin{figure}[ht]
\centering
\includegraphics[width=.5\linewidth]{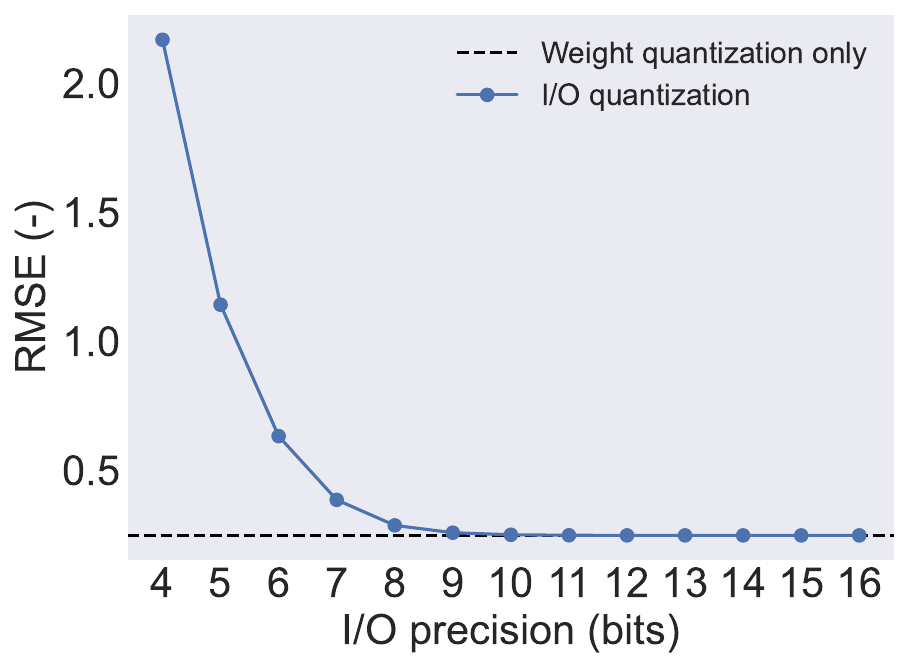}
\caption{RMSE in randomized MVM results as a function of the input/output quantization precision $b$ (bits), compared to the weight-quantization-only floor (dashed line) for $N =$4 states.}\label{fig:fig_io_quant}
\end{figure}

The results are shown in Figure~\ref{fig:fig_io_quant}.
At low precision ($b \leq 6$ bits), input/output quantization dominates the error and the RMSE significantly exceeds the weight-quantization floor (1/4).
Above $b \approx 10$ bits, however, the RMSE converges to the weight-quantization-only value, meaning that the interface quantization becomes negligible.
Standard laboratory wave generators and oscilloscopes operate at 12 to 16 bit precision, placing them well above this threshold.
Hence, for the experimental validation scope of this work, input/output quantization does not constitute a significant additional error source, and weight quantization remains the dominant contributor to MVM error.
However, for chip-integrated implementations where ADC/DAC precision is limited to 4 to 6 bits, interface quantization would represent a non-negligible additional error term that would need to be accounted for.

\subsubsection*{Systematic errors}
While quantization error is inherent to the hardware inference pipeline, external sources of error also have an impact on the accuracy of the MVMs performed with the crossbar array. 
First, we investigate systematic cell non-idealities whose origin is common to all the cells in the crossbar array.
These non-idealities induce a shared bias in the properties of the cells, and may lead to significant performance degradation.
Although the array-aware quantization calibrates the states to the array's mean resistance levels, a given array may deviate from this reference: process variations affecting all cells of an array similarly, or a global drift due to temperature or ageing, displace the whole level set away from the calibrated values.
Such a deviation is not removed by the calibration and constitutes a systematic non-ideality as it is shared by all the array cells.
In practice, we first consider that all the cells originally access the same reference set of resistance levels.
We then model the non-idealities by adding a random offset drawn from $\mathcal{N}(0, \sigma_\text{NL})$ to each resistance level.
The resulting deviation is shared by all the cells of the crossbar array to ensure systemacy.
We then compute the RMSE in the result of randomized MVMs in a simulated array of (4 $\times$ 4) cells presenting 4 resistance states like previously, for different values of $\sigma_\text{NL}$.
Figure~\ref{fig:fig_rmse_comp} shows that the RMSE in the MVM result increases linearly with $\sigma_\text{NL}$. 
However, the RMSE is not equal to 0 in the absence of non-idealities ($\sigma_\text{NL}=$ 0) due to the presence of the quantization error. 
Indeed, at low $\sigma_\text{NL}$ values, the resistance level distribution remains close to the reference and the RMSE is dominated by the quantization contribution (1/4), marking a lower limit of the RMSE in this case.

\subsubsection*{Cell-specific errors}
\begin{figure}[ht]
\centering
\includegraphics[width=.5\linewidth]{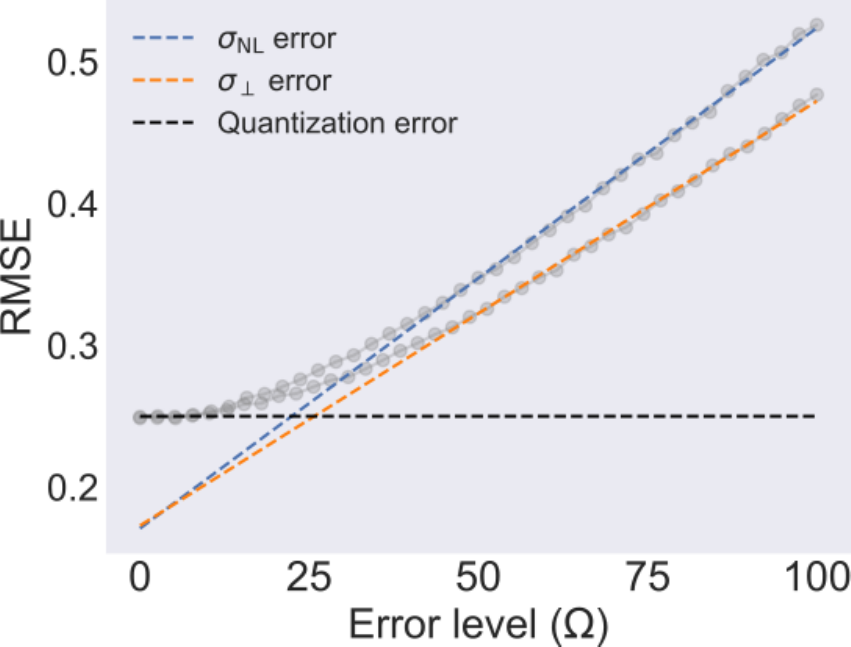}
\caption{RMSE in randomized MVM results with respect to the level of the error source in a ($4 \times 4$) crossbar array of $4$-states M²TJs.}\label{fig:fig_rmse_comp}
\end{figure}

We then investigated the presence of non-idealities specific to each cell, due for example to sources such as cell-to-cell variations and noise.
To simplify the study, we consider a combination of all of these non-idealities by adding a second random offset to each resistance state drawn from a normal distribution $\mathcal{N}(0, \sigma_\perp)$.
The key difference with the previously introduced systematic non-idealities is that these random offsets are independent between each cell, and can hence benefit from averaging effects over the whole array.
We repeated the same study as for systematic non-idealities, and observed that while the RMSE measured in random MVM results also evolves linearly with $\sigma_\perp$ as seen in Figure~\ref{fig:fig_rmse_comp}, the increase in RMSE is slower with $\sigma_\perp$ than with $\sigma_\text{NL}$ thanks to mitigation by averaging.
The Gaussian distributions used here represent an approximation of device variability.
The framework itself does not impose any choice of distribution: replacing the Gaussian model with any measured or empirical distribution (for example to account for non-Gaussian tails) requires only substituting the distribution used in the simulation, without any modification to the retrieval formula.
Using realistic noise distributions characterized from experimental data is an important direction for future work.

\subsubsection*{Optimal number of states per cell}
\begin{figure}[ht]
\centering
\includegraphics[width=\linewidth]{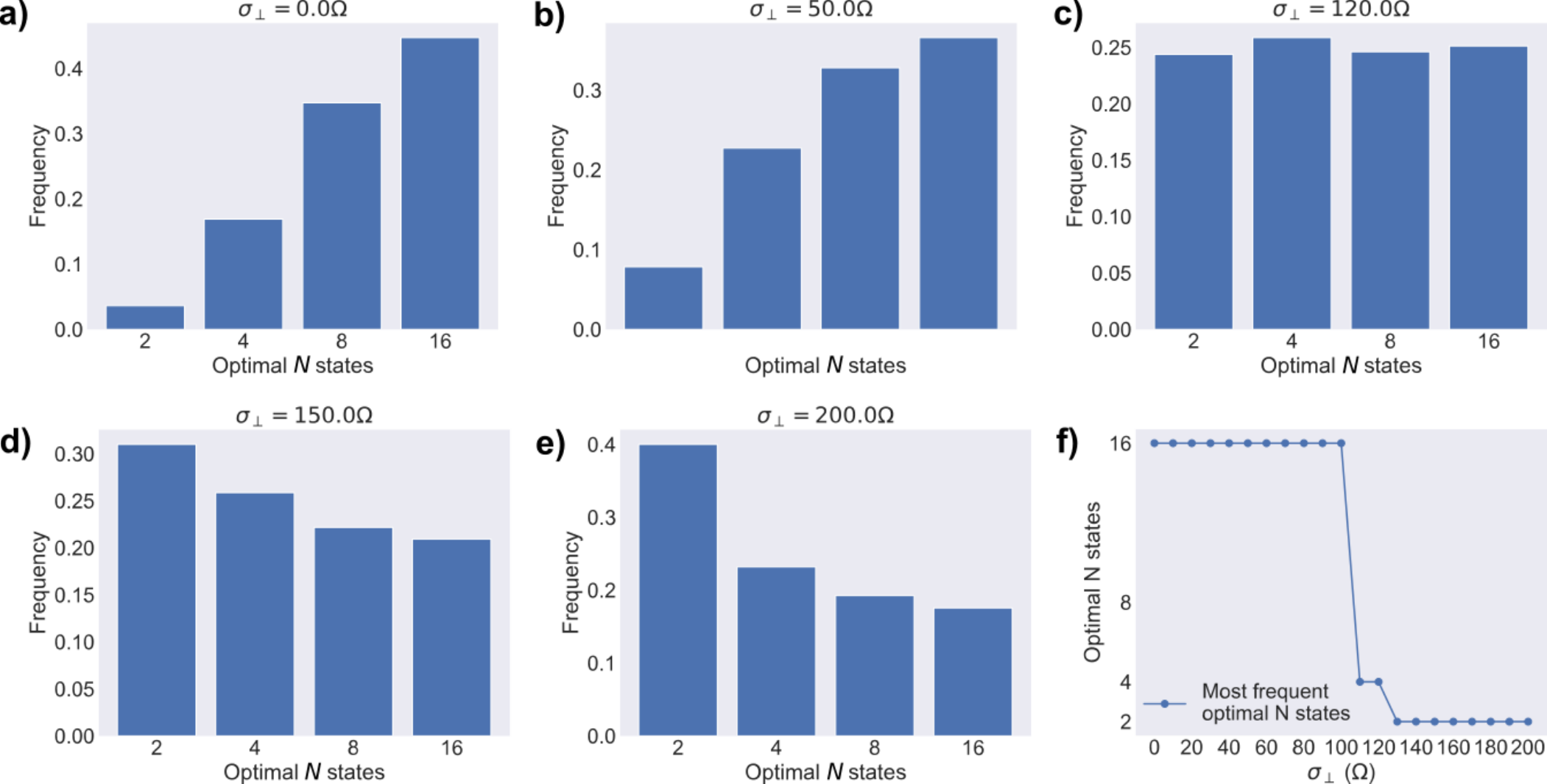}
\caption{\textbf{a-e}) Normalized distribution of the optimal number of states $N_\text{opt}$ minimizing the RMSE in random MVM results, for $\sigma_\text{NL} =$ 50 $\Omega$ and increasing $\sigma_\perp$ values. \textbf{f)} Optimal number of states with respect to the level of cell-specific noise $\sigma_\perp$}\label{fig:fig_nopt}
\end{figure}

Finally, we investigated the impact of the number of states per cell $N$ on the accuracy of the MVM results in the presence of non-idealities.
Indeed, while increasing $N$ allows to reduce the quantization error, it also decreases the resolution between the resistance states, which can be detrimental in the presence of non-idealities.
The optimal number of states per cell $N_\text{opt}$ is thus a trade-off between the minimization of the quantization error and the maximization of the states resolution, which is itself proportional to the total range of accessible resistance values $[R_\text{min}, R_\text{max}]$ and inversely proportional to the effective non-idealities deviation $\sigma_\text{tot} = \sqrt{\sigma_\text{NL}^2+\sigma_\perp^2}$. 
The normalized distribution of the value of $N$ minimizing the RMSE in random MAC operations results for $\sigma_\text{NL} = 50$ $\Omega$ and $\sigma_\perp$ values from 0 $\Omega$ to 200 $\Omega$ is shown in Figure~\ref{fig:fig_nopt}.
In low-noise environments, the quantization error is predominant and higher $N_\text{opt}$ values allow to reduce the RMSE.\@
As $\sigma_\perp$ increases and exceeds 100 $\Omega$, $N_\text{opt}$ quickly drops due to the loss of resolution between successive resistance levels, confirming the predictions.
It is however worth noting that $N_\text{opt}$ starts decreasing when $\sigma_\perp$ exceeds 100 $\Omega$, which is a significant deviation compared to the distance between two successive states (181 $\Omega$), highlighting the robustness of the MVM results to non-idealities in this case. 
Quantization error is hence the dominant source of error in this case, and the use of multistate devices with $N$ higher than 4 would be beneficial to improve the accuracy of the MVM results.
Figure~\ref{fig:fig_nopt}f) shows the evolution of $N_\text{opt}$ with respect to $\sigma_\perp$.
The step-like appearance of the curve is due to the fact that $N_\text{opt}$ is an integer value constrained in the [2, 16] range.

\subsubsection*{Error scaling}
We finally consider a crossbar array with fixed levels of systematic and cell-specific non-idealities $\sigma_\text{NL}$ and $\sigma_\perp$, and we investigate the scaling of the RMSE in the MVM results with respect to the size of the array. We assume that this scaling is only due to the increase of the number of cells in the array, and that the RMSE scales similarly in a multiplexed array of small size and in a geometrically larger array.

\begin{figure}[ht]
\centering
\includegraphics[width=.8\linewidth]{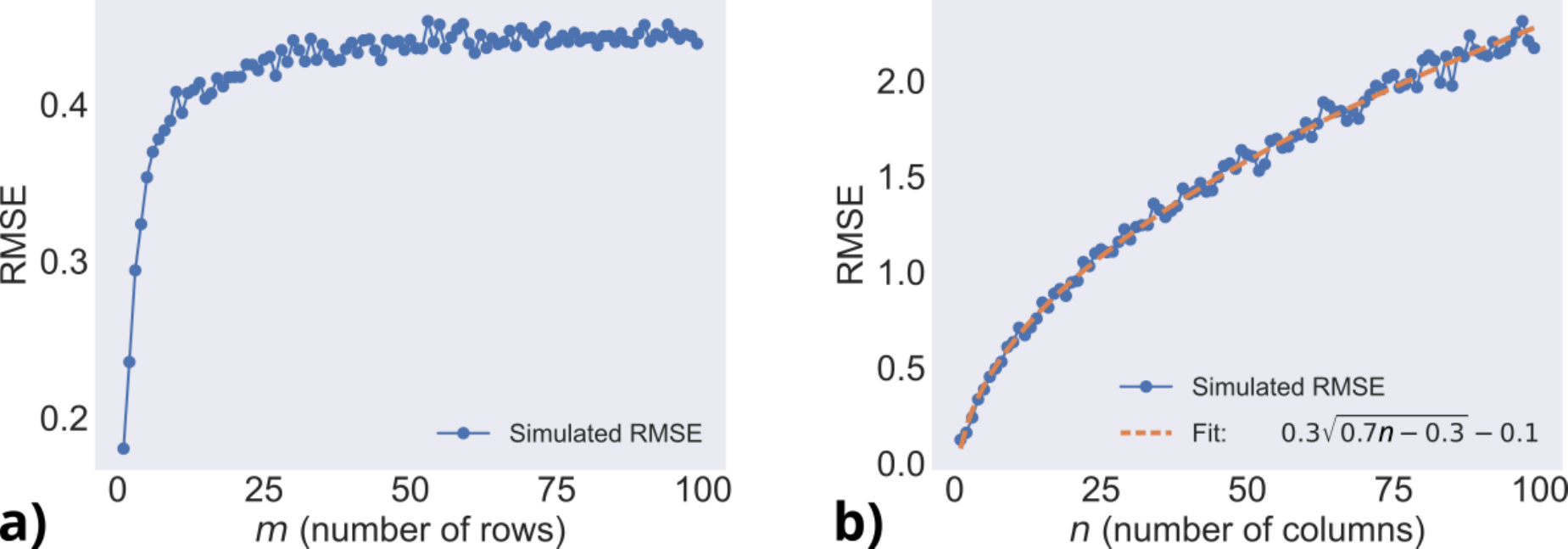}
\caption{\textbf{a)} Scaling of the RMSE in MVM results with the number of rows $m$ in the matrix $\textbf{A}$, for $\sigma_\text{NL} =$ 0 $\Omega$ and $\sigma_\perp =$ 50 $\Omega$. \textbf{b)} Scaling of the RMSE in MVM results with the number of columns $n$ in the matrix $\textbf{A}$, for $\sigma_\text{NL} =$ 0 $\Omega$ and $\sigma_\perp =$ 50 $\Omega$.}\label{fig:fig_rmse_scaling}
\end{figure}

\begin{figure}[ht]
\centering
\includegraphics[width=.8\linewidth]{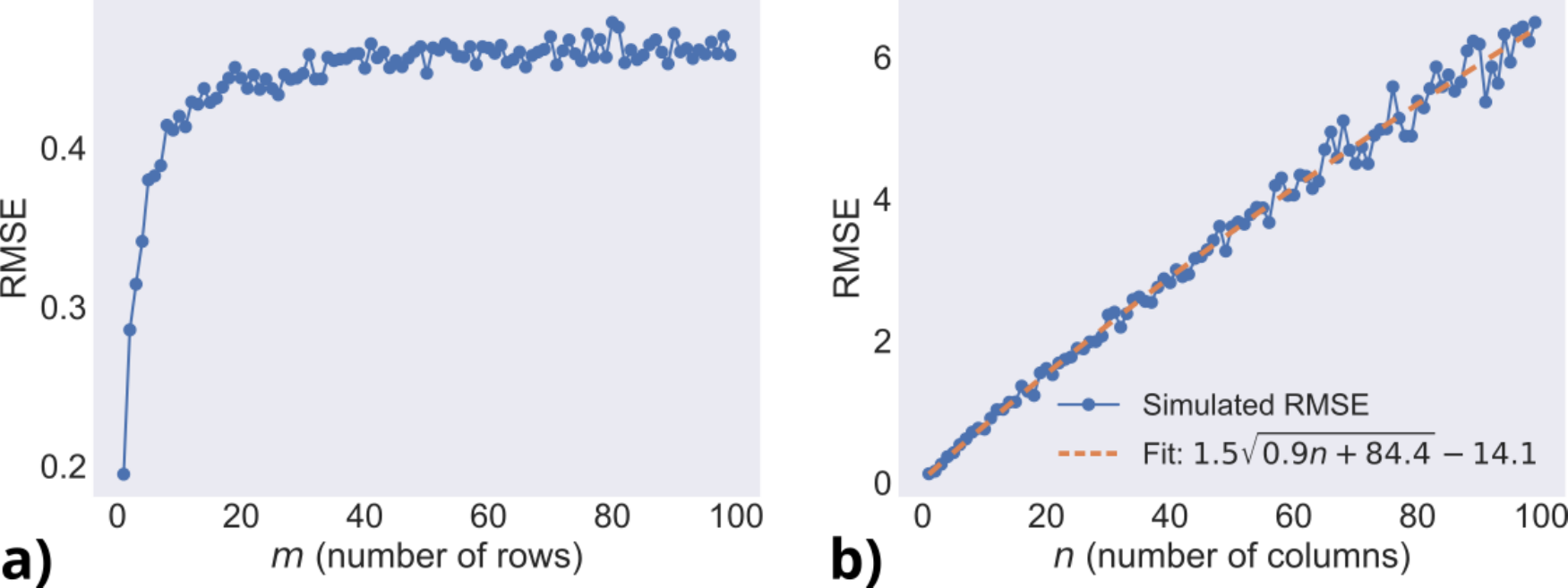}
\caption{\textbf{a)} Scaling of the RMSE in MVM results with the number of rows $m$ in the matrix $\textbf{A}$, for $\sigma_\text{NL} =$ 50 $\Omega$ and $\sigma_\perp =$ 0 $\Omega$. \textbf{b)} Scaling of the RMSE in MVM results with the number of columns $n$ in the matrix $\textbf{A}$, for $\sigma_\text{NL} =$ 50 $\Omega$ and $\sigma_\perp =$ 0 $\Omega$.}\label{fig:fig_rmse_scaling2}
\end{figure}

When $\sigma_\text{NL} =$ 0 $\Omega$ and $\sigma_\perp =$ 50 $\Omega$ (\textit{i.e.}, the cells are ideal but subject to noise), we observe that the total RMSE in the MVM results is constant with the number of rows $m$ in the matrix $\textbf{A}$ (which is either related to the size of the crossbar array or to the number of multiplexing steps required), after a short transient regime at low $m$ values as seen in Figure~\ref{fig:fig_rmse_scaling}~a. This is due to the fact that rows correspond to independent output channels of the crossbar array, and that the error in each output channel is drawn from the same distribution. 
Hence, adding a new row in $\textbf{A}$ simply adds a new sample of the same error distribution in the output vector, leaving the RMSE unchanged.
However, it can be seen that the RMSE increases with the square root of the number of columns $n$ in the matrix $\textbf{A}$ (Figure~\ref{fig:fig_rmse_scaling}~b), because the noise-induced error of the columns effectively accumulates along each row.
When $\sigma_\text{NL} =$ 50 $\Omega$ and $\sigma_\perp =$ 0 $\Omega$ (\textit{i.e.}, the cells are subject to systematic non-idealities but no noise), the RMSE in the MVM results scales similarly as in the previous case \textit{i.e.}, constant with the number of rows $m$ and increasing with the square root of the number of columns $n$ (Figure~\ref{fig:fig_rmse_scaling2}). However, the increase of the RMSE with $n$ is more significant in this case, which can be explained here again by the fact that systematic non-idealities are not mitigated by averaging effects over the whole array as opposed to cell-specific non-idealities.

\section*{Discussion}
The evaluation on the XOR approximation and MNIST classification tasks confirms that the crossbar array inference process achieves good agreement with the software baseline in both cases, with as few as 4 states per cell. 
This highlights the relevance of multistate memristive devices for IMC, even with a moderate number of states, and is supported by the clear $1/N$ dependence of the quantization error on the number of states per cell $N$.
However, the latter conclusion must be balanced against the impact of device non-idealities, which constitute the two other error sources investigated in this work.
Systematic non-idealities, which introduce a shared bias across cell properties, and cell-specific non-idealities, which affect each cell independently, cause the RMSE in the MVM results to increase linearly with the deviation from the ideal linear state distribution.
The increase is slower for cell-specific non-idealities thanks to mitigation through averaging effects across the array.
The distinction between both sources of error also shows in the scaling of the RMSE with the size of the array. 
The RMSE remains constant with the number of rows $m$ in both cases, but grows with the square root of the number of columns $n$, with a more pronounced increase for systematic non-idealities.
These results suggest that array geometry and the nature of the dominant source of error should both be considered when optimizing hardware configurations.
While the trade-off between quantization error and states resolution leads to the existence of an optimal number of states $N_\text{opt}$, the MVM results demonstrate robustness to non-idealities, meaning that $N$ values well above 4 can still be beneficial even under significant noise levels.
This observation gives confidence to the use of memristive devices with a high number of states in real inference processes.
The framework presented here provides a general methodology for predicting the performance of crossbar array-based inference processors from the physical parameters of the memristive cells.
By characterizing the actual noise and non-ideality levels of a given hardware solution, the framework also enables data-driven determination of the optimal $N$ to minimize MAC operation error.

Sneak currents, who arise from the flow of current though unselected cells, and parasitic resistance, who introduce detrimental voltage (IR) drops along the rows and columns of the array, are two additional sources of systematic errors ($\sigma_\text{NL}$) that can be witnessed in passive crossbar arrays.
However, they were not included in the error studies of this work, as they are highly dependent on the specific architecture of the crossbar array and its integration, and require explicit circuit-level modeling to be properly accounted for.
Incorporating explicit circuit-level models of sneak currents and IR drop into the framework is thus an important direction for future works.
In the same way, a system-level energy and latency analysis, which requires specifying the device technology, peripheral circuit architecture, and operating conditions, is beyond the scope of this work but is a natural extension once the framework is applied to a specific hardware implementation.

The simulation parameters used in this work are derived from experimentally characterized M²TJs reported in Refs.~\cite{Das2020, Das2020stabilization}, providing a direct link between the framework and measured device data.
The framework is designed as a tool to guide the design before fabrication, as it is meant to predict the expected inference accuracy before a crossbar array is built, allowing informed design choices such as the selection of the number of resistance states $N$ and the impact of the expected non-idealities.
Experimental validation of crossbar inference against the framework's predictions is planned as part of the ongoing MultiSpin.AI project and will be reported in future works.

\section*{Methods}
\subsection*{The N-ary MAC operation}
We will use Eq.~\ref{eq:mac_eq1bis} and Eq.~\ref{eq:mac_eq2bis} to implement Eq.~\ref{eq:mvm_eq} in the memory. 
To do so, we encode the full-precision weight matrix $\mathbf{W}$ into a matrix $\mathbf{A}$ whose entries belong to a set of $N$ distinct values representing the distinct resistance levels of the crossbar array cells.
This allows to Eq.~\ref{eq:mac_eq3} as an approximation of the result of Eq.~\ref{eq:mvm_eq}
\begin{equation}
    \tilde{\mathbf{y}} = \mathbf{A} \mathbf{x}
    \label{eq:mac_eq3}
\end{equation}

Despite the apparent simplicity of the operation, a challenge is to retrieve $\tilde{\mathbf{y}}$ based solely on $\mathbf{A}$, $\mathbf{x}$, and the measured voltage vector $\mathbf{V}$ containing the output values $V_m$.
While some studies circumvent this step by including the crossbar array physics directly within the network's training procedure, we aim at using the hardware as a standalone step in the inference pipeline to increase modularity and generalizability to other AI frameworks. 
To do so, we first assume that the input $\mathbf{x}$ is linearly encoded into a current intensity vector $\mathbf{I}$ suiting the cell devices physics (Eq.~\ref{eq:eq_intensity}).
\begin{equation}
    \mathbf{I} = a_I \mathbf{x} + b_I
    \label{eq:eq_intensity}
\end{equation}
We similarly \textit{require} that the $N$ resistance levels of the cells can be expressed as a linear function of the values of matrix $\mathbf{A}$ (Eq.~\ref{eq:r_levels}).
This does not assume that the physical levels are evenly spaced as detailed in the Weights quantization section.
In fact, the values $A_i$ are assigned with the relative spacing of the measured resistance levels, so that Eq.~\ref{eq:r_levels} is exact for any given set of levels.
\begin{equation}
    R_{m n} = a_R A_{m n} + b_R 
    \label{eq:r_levels}
\end{equation}
Owing to Eqs.~\ref{eq:mac_eq1bis} and \ref{eq:mac_eq2bis}, we can write
\begin{align}
   V_m &= \sum_n (a_R A_{m n} + b_R) (a_I x_n + b_I)\\
   &= \sum_n (a_R a_I A_{m n} x_n + b_R a_I x_n + a_R b_I A_{m n} + b_R b_I)\\
   &= a_R a_I \sum_n A_{m n}x_n + b_R a_I \sum_n x_n + a_R b_I \sum_n A_{m n} + n b_R b_I
\end{align}
And as $\tilde{y}_m = \sum_n A_{m n} x_n$, one can write 
\begin{equation}
    \tilde{y}_m = (V_m - b_R a_I \sum_n x_n - a_R b_I \sum_n A_{m n} - n b_R b_I ) / (a_R a_I)
    \label{eq:VB}
\end{equation} which under vectorial form writes
\begin{equation}
    \tilde{\mathbf{y}} = (\mathbf{V} - b_R (n b_I + a_I (\mathbf{x}^T \cdot \mathbf{1})\mathbf{1}-a_R b_I (\mathbf{A} \cdot \mathbf{1}))) / (a_R a_I)
    \label{eq:VBvec}
\end{equation}
This equation can be further simplified by replacing the $(n b_I + a_I (\mathbf{x}^T \cdot \mathbf{1}))$ term by $(\mathbf{I}^T \cdot \mathbf{1})$, yielding
\begin{equation}
    \tilde{\textbf{y}} = (\textbf{V} - b_R (\textbf{I}^T \cdot \textbf{1})-a_R b_I (\textbf{A} \cdot \textbf{1})) / (a_R a_I
    \label{eq:eq_VB}
\end{equation}
where $a_R$, $a_I$, $b_R$ and $b_I$ are known constants.
The term $(\textbf{I}^T \cdot \textbf{1})$ corresponds to the sum of the input intensities, while $(\textbf{A} \cdot \textbf{1})$ is the column-wise sum of matrix $\textbf{A}$, which needs to be computed only once and can then be stored in adjacent memory.
The post-processing step that implements Eq.~\ref{eq:eq_VB} involves two additional quantities: the sum of input intensities $(\textbf{I}^T \cdot \textbf{1})$, which can be computed digitally from the known input vector before the MVM, and the column-wise sum $(\textbf{A} \cdot \textbf{1})$, computed only once after weight programming and stored in adjacent digital memory.
While both operations are purely digital, they add negligible overhead compared to the analog MVM itself.
This post-processing step can be implemented in peripheral circuitry close to the crossbar or directly in the measurement equipment, leaving the internal architecture of the array untouched.
A more detailed study of the actual hardware architecture and the digital peripheral circuitry will be carried out in the upcoming stages of the project. 

\subsection*{Weights quantization}
The weights matrix $\textbf{W}$ learned during the training of the neural network must be quantized into a matrix $\textbf{A}$ whose elements belong to a set of $N$ values representing the $N$ distinct conductance (or resistance) levels of the crossbar array cells.
As the framework is \textit{array-aware}, the $N$ states $A_i$ are not forced to be equidistant. 
Instead, their relative spacings are set equal to those of the device's measured resistance levels $R_i$, so that $A_i - A_{i-1} = \rho_i\, d$ with $\rho_i = (R_i - R_{i-1})/(R_N - R_1)$ the normalized spacing of the measured levels and $d$ a common factor.
This guarantees that the state-to-resistance mapping of Eq.~\ref{eq:r_levels} is exactly linear.
The states $A_i$ must then minimize the sum of the squared distances between each weight $w$ and the closest value $A_i$, subject to this spacing constraint.
This problem, formalized in Eq.~\ref{eq:opti}, is similar to the $k$-means problem~\cite{Macqueen1965}, except that in our case the inter-state relative distances $\rho_i$ are fixed by the measured resistance spacings.
We determine the optimal value of the first state $A_1$ and that of the optimal factor $d$ using the \texttt{minimize} function from the Scipy Python library and the Powell minimization method~\cite{Fletcher1963}.
The optimization operates only on parameters $A_1$ and $d$, regardless of the number of weights in the layer, which only constitute the dataset over which the cost function is evaluated.
\begin{equation}
      \text{argmin}_{A} \sum_{i=1}^N \sum_{w \in A_i} |w-A_i|^2 \quad \text{with} \quad A_i - A_{i-1} = \rho_i\, d
      \label{eq:opti}
\end{equation}
In the special case of an equidistant device ($\rho_i$ constant for all $i$), the standard equidistant quantization is recovered.

\bibliography{sample}

\section*{Funding}
This work is part of the MultiSpin.AI project. MultiSpin.AI has received funding from EU under grant agreement number 101130046. Views and opinions expressed are however those of the author(s) only and do not necessarily reflect those of the European Union or the European Innovation Council and SMEs Executive Agency (EISMEA). Neither the European Union nor the granting authority can be held responsible for them.

\section*{Author contributions statement}
A. M. and A. L. T. conceived the study, carried out the simulations, and wrote the manuscript.
F. A. A. supervised the study and contributed to the analysis of the results. 

\section*{Code availability}
The code used to carry out the simulations and generate the pictures of this manuscript can be asked to the authors upon reasonable request.

\section*{Data availability}
The simulation of the multistate MTJs properties were based on the data reported in Refs.~\cite{Das2020, Das2020stabilization}. The dataset used for the MNIST classification task is publicly available from the Tensorflow.Keras Python library.

\section*{Competing interests}
The authors declare no competing interests.

\end{document}